\documentstyle[aps, eqsecnum,preprint]{revtex} \baselineskip 12pt
\begin{document}
\title{ On an exactly solvable $B_N$ type Calogero model \\
          with nonhermitian PT invariant interaction } 

\vspace{.4in}

\author{B. Basu-Mallick }

\address{ Theory Group, Saha Institute of Nuclear Physics,\\
1/AF-Bidhannagar, Calcutta-700064, India.\\
Email: biru@tnp.saha.ernet.in
}

\author{Bhabani Prasad Mandal}
\address{
S.N. Bose National Center for Basic Sciences,\\
Block-JD; Sector-III; Salt Lake ,\\
Calcutta-700 098, India.\\
Email:  bpm@boson.bose.res.in 
}

\vspace{.4in}

\maketitle

\vspace{.4in}

\begin{abstract}

An exactly solvable many-particle
quantum system is proposed by adding some nonhermitian 
but PT invariant interactions to the $B_N$ Calogero model.
 We have shown that such extended 
 $B_N$ Calogero model leads to a real spectrum 
 which obey generalised exclusion statistics.  It is also found that
 the corresponding exchange statistics parameter exhibit  
`reflection symmetry' provided the strength of a PT invariant interaction 
 exceeds a critical value.

\end{abstract}

\newpage

\section{Introduction} 
As is well known, in quantum mechanics one usually chooses a hermitian 
Hamiltonian to ensure real energy eigenvalues of the corresponding 
Schr${\rm {\ddot o}}$dinger equation.
However, quantum mechanical systems characterised by 
nonhermitian Hamiltonians also play a significant role in many contexts 
like absorption of incident particles in nuclear physics, 
localisation-delocalisation transitions in superconductors and in 
the description of the defraction of atoms by standing 
light waves \cite {ber,ber1}. Recently,  theoretical investigations on
different nonhermitian Hamiltonians have received a major
 boost due to the remarkable observation that many such  systems, 
whenever they are invariant under combined parity (P)
and time reversal (T)  symmetry,  give real  energy 
eigenvalues \cite {bd}.
 This seems to suggest that the condition of hermiticity 
on a Hamiltonian can be replaced by the weaker condition of PT symmetry 
to ensure that the corresponding  eigenvalues would be real ones. 
However, till now this is merely a conjecture supported by several examples
\cite {bd,oth,bd1,bd2,bd3,bd4,bd5,bd6,kb,bd7,bd8,bd9,bd10}. 
Moreover, in almost all of these examples, the Hamiltonians of 
only one particle in one space dimension have been considered.  

The aim of the present work is to test the validity of
 the above mentioned conjecture 
for the cases of some exactly solvable  many-particle quantum mechanical
systems in one dimension. For a Hamiltonian
 containing $N$ number of particles,  
   the $PT$ transformation is evidently given by 
\begin{equation} 
i \rightarrow -i, \ \ x_j \rightarrow - x_j, \ \  p_j \rightarrow p_j 
\label {pt}
\end{equation} 
where $j \in [1,2, \cdots , N],$  and 
$x_j$ ($p_j \equiv -i \frac {\partial }{\partial  x_j}$) denote the coordinate 
(momentum) operator of the $j$-th particle. 
So we want to find out 
 some exactly solvable nonhermitian Hamiltonians which remain
invariant under the PT transformation (\ref {pt}) and 
  investigate whether such systems would lead to 
 real spectra.  

In this context one may note that the well known $A_{N-1}$ 
Calogero model,  which contains $N$ particles on a line and is 
 described by the hermitian Hamiltonian
\begin{equation} 
H_A= - \frac{ 1}{2} \sum_{j=1}^N \frac{ \partial ^2}{\partial x_j^2} +
\frac{1}{2} \omega ^2 \sum_{j=1}^Nx_j^2 + 
 +{g\over 2 } \sum_{\stackrel{j,k =1}{(j\neq k)}}^N
 \frac{1}{(x_j-x_k)^2} \ ,
\label{cs}
\end{equation} 
represents an exactly solvable system \cite {cm}.
This type of exactly solvable models with long-range 
interaction have attracted a lot of 
attention due to their close connection with diverse subjects like 
fractional statistics, random matrix theory, level statistics 
for disordered systems etc. \cite {app1,app2,app3,app4,app5,ms,ms1,ms2}. 
Recently an integrable extension of $A_{N-1}$ Calogero model
 is proposed by adding some momentum dependent 
interaction 
 to the Hamiltonian (\ref {cs}) and this extension of Calogero model 
 is also solved exactly to obtain the 
energy eigenvalues as well as eigenfunctions \cite {ab}. 
It is found that
these energy eigenvalues are real and bounded below, in spite of the 
fact that the added momentum dependent interaction given by
\begin{equation} 
H_p = {\delta \over 2} \sum_{\stackrel {j,k=1}{j\neq k}}^N
 {1\over x_j -x_k } 
\left ( {\partial  \over \partial  x_j} - {\partial  \over \partial  x_k} 
\right ) \  , \label{hp}
\end{equation} 
where $\delta $ is a real parameter, is not a hermitian operator.

However, we now interestingly observe 
 that the complete Hamiltonian ($H_A +H_p)$ of 
the above mentioned extended Calogero model is indeed invariant 
under PT transformation (\ref{pt}). Thus we find here an example of 
PT invariant many-particle system, which is not only exactly solvable 
but also leads to completely real spectrum. 
To obtain more examples of this type,
 it may be noted that the Calogero 
model associated with $B_N$ root system can  also be solved exactly 
\cite {bn,bn1}.  Therefore, it is natural to
ask whether this $B_N$ Calogero model can be generalised in a 
 PT invariant way so that the newly constructed model
 would remain exactly solvable 
and yield completely real spectrum. 
In Sec.II we try to answer this question by proposing an appropriate
 PT invariant
 extension of the $B_N$ Calogero model and solving such model exactly. 
In Sec.III we study some salient features of this  extended $B_N$ 
Calogero model and show that its spectrum can be interpreted through 
 generalised exclusion statistics \cite {hal} as 
proposed by Haldane. Sec.IV is the concluding section.

\vspace{.6in}

\section{Exact spectrum of a $B_N $ type Calogero model}

Here we propose an extension of 
 the well known $B_N$ Calogero model \cite {bn,bn1} as 
\begin{eqnarray}
{\cal H}_B &=& - 
\frac{ 1}{2} \sum_{j=1}^N \frac{ \partial ^2}{\partial x_j^2} +
\frac{ 1}{2} \omega ^2 \sum_{j=1}^Nx_j^2 + \frac{ g_1}{2}\sum_{j=1}^N
\frac{ 1}{x_j^2} +g_2\sum_{\stackrel{j,k=1}{j\neq k}}^N \frac{ x_j^2
+x_k^2}{(x_j^2-x_k^2)^2}\nonumber \\ 
&+& \delta _1\sum_{j=1}^N \frac{ 1}{x_j} \frac{ \partial }{\partial x_j} +
\delta _2\sum_{\stackrel{j,k=1}{j\neq k}}^N
 \frac{ 1}{x_j^2 -x_k^2 }\left (x_j \frac{ \partial
}{\partial x_j} - x_k \frac{ \partial }{\partial x_k} \right ),
\label{ha}
\end{eqnarray}
where $ g_1,\ g_2,\ \delta _1,\ \delta _2 $ are some
 real coupling constants.  In the special case
 $ \delta _1= \delta _2 =0$,  the Hamiltonian 
 (\ref{ha}) reproduces the original $B_N$ Calogero model:
\begin{equation}
H_B = - \frac{ 1}{2} \sum_{j=1}^N \frac{ \partial ^2}{\partial x_j^2} +
\frac{ 1}{2} \omega ^2 \sum_{j=1}^Nx_j^2 + \frac{ g_1}{2}\sum_{j=1}^N
\frac{ 1}{x_j^2} +g_2\sum_{\stackrel{j,k=1}{j\neq k}}^N 
 \frac{ x_j^2 +x_k^2}{(x_j^2-x_k^2)^2} .
\label{hb}
\end{equation}
It may be observed that, though the Hamiltonian (\ref {ha})
violates hermiticity property due to the presence of 
momentum dependent interactions like 
$\delta _1\sum_{j=1}^N \frac{ 1}{x_j} \frac{ \partial }{\partial x_j} +
\delta _2\sum_{j\neq k} \frac{ 1}{x_j^2 -x_k^2 }\left (x_j \frac{ \partial
}{\partial x_j} - x_k \frac{ \partial }{\partial x_k} \right )$,
 it remains invariant under the
 combined PT transformation (\ref {pt}). Therefore,
it should be interesting to solve  this 
 extended $B_N $  Calogero model and see whether it gives real 
energy eigenvalues.

 In this context it may be noted that,
 both $A_{N-1}$ and $B_N$ Calogero  models
have been solved recently by mapping them  to 
a system of free  harmonic oscillators \cite{osc,gp,sog,bak,wad0,wad}. 
For the purpose of solving 
 the Hamiltonian (\ref {ha}) through 
 a similar procedure, 
we conjecture first that its ground state is given by a
 Laughlin type wave function 
\begin{equation}
\psi_{gr} = \prod_{\stackrel{j,k =1}{(j\neq k)}}^N (x_j -x_k)^\rho
(x_j+x_k)^\rho \prod_{l=1}^Nx_l^\sigma \ e^{- \frac{ \omega }{2}\sum_{j=1}^N
x_j^2} ,
 \label{gr}
\end{equation}
  $ \rho$ and $\sigma $ being  two  real 
non-negative parameters which are related
  to the  coupling constants $ g_1, g_2, \delta _1, \delta _2 $ as
\begin{equation}
g_1 = \sigma (\sigma -1) - 2 \sigma
 \delta _1   \ ; \ \ \ \ \  g_2 = 2 \rho (
2\rho -1) - 4\rho \delta _2 .
\label{g1g2}
\end{equation}
By solving the quadratic equations (\ref {g1g2}) of 
 ${\sigma }$ and ${\rho} $,  it is easy to see that these two parameters
  take real values provided 
the coupling constants in the Hamiltonian (\ref{ha}) satisfy 
\begin{equation}
g_1\ge -( \frac{ 1}{2} + \delta _1)^2 \ ; \ \ \  g_2 \ge -( \frac{ 1}{2}+
\delta _2)^2.
\label{cond}
\end{equation}
So, in this article we shall consider the PT invariant Hamiltonian 
(\ref{ha}) only for the range of parameters compatible with the conditions
(\ref {cond}).  Now if we 
 use the expression (\ref{gr})  for a similarity 
transformation to the Hamiltonian (\ref {ha}), it reduces 
 to a simple `effective Hamiltonian' like 
\begin{eqnarray}
\tilde {\cal H}_B &=& 
\psi_{gr}^{-1} \left ( {\cal H}_B - E_{gr} \right ) \psi_{gr} \nonumber
\\ 
&=& - \frac{ 1}{2}\sum_{j=1} ^N \frac{ \partial ^2}{\partial x_j^2} +
\omega \sum_{j=1}^N x_j \frac{ \partial }{\partial x_j} -2\tilde{\rho}
\sum_{\stackrel {j,k =1}{j\neq k}}^N 
\frac{ 1}{x_j^2 - x_k^2} \left ( x_j \frac{ \partial
}{\partial x_j} - x_k \frac{ \partial }{\partial x_k} \right ) - \tilde{
\sigma }\sum_{j=1}^N \frac{ 1}{x_j} \frac{ \partial }{\partial x_j} .
\label{h1}
\end{eqnarray} 
Here
\begin{equation}
E_{gr} = \frac{ \omega N}{2} + \omega N \tilde{ \sigma }+ \omega
N(N-1)2\tilde{\rho},
\label{egr}
\end{equation}
and the real valued parameters
 $\tilde{\sigma }, \ 2\tilde{\rho} $ are defined as
 $\tilde{\sigma } = \sigma - \delta _1$ and $ 2\tilde{\rho} =
2\rho - \delta _2$ .

It should be noted that, at the limit $\delta_1 = \delta_2 = 0$, 
$\tilde {\cal H}_B $ (\ref {h1}) reduces to 
\begin{equation}
\tilde H_B = 
 - \frac{ 1}{2}\sum_{j=1} ^N \frac{ \partial ^2}{\partial x_j^2} +
\omega \sum_{j=1}^N x_j \frac{ \partial }{\partial x_j} -2{\rho}
\sum_{\stackrel {j,k =1}{j\neq k}}^N 
\frac{ 1}{x_j^2 - x_k^2} \left ( x_j \frac{ \partial
}{\partial x_j} - x_k \frac{ \partial }{\partial x_k} \right ) - {
\sigma }\sum_{j=1}^N \frac{ 1}{x_j} \frac{ \partial }{\partial x_j},
\label{h11}
\end{equation} 
where the parameters $ \rho$ and $\sigma $ are related with 
  the coupling constants $ g_1, g_2$ of the $B_N$ Calogero model
(\ref {hb}) as
\begin{equation}
g_1 = \sigma (\sigma -1) \ ; \ \ \ \  g_2 = 2 \rho ( 2\rho -1) .
\label{gg}
\end{equation}
So, by
 using the expression (\ref{gr})  for a similarity transformation 
on the $B_N$ Calogero model
 (\ref {hb}) and applying the relations (\ref {gg}), one can get 
the corresponding effective Hamiltonian (\ref {h11}). 
 It has been found earlier that this effective Hamiltonian  (\ref {h11})
 has a complete set of polynomial 
eigenfunctions \cite {bak1}. 
Such polynomial eigenfunctions, which are completely 
symmetric with respect to variables $x_i^2$, are called as generalised 
Laguerre polynomials. 
Since these  symmetric polynomial eigenfunctions 
are evidently free from any singularity,
  there exists an one-to-one 
correspondence between the nonsingular eigenfunctions of the 
   usual $B_N$ Calogero Hamiltonian and the generalised 
Laguerre polynomials.  Now it is very interesting 
to observe that, 
after a trivial substitution of coupling constants given by 
$ 2\tilde{\rho} \rightarrow 2\rho $ and
$\tilde{\sigma } \rightarrow \sigma $,
the  effective Hamiltonian (\ref {h1}) for 
our extended $B_N$ Calogero model coincides with the
 effective Hamiltonian (\ref {h1})  for 
 $B_N$ Calogero model.  Therefore, 
the  generalised Laguerre polynomials again represent
 a complete set of polynomial eigenfunctions  for 
 the effective Hamiltonian  (\ref {h1}). Consequently, 
 there exists  an one-to-one correspondence between these 
 polynomials and the eigenfunctions of
 the extended $B_N$ Calogero Hamiltonian (\ref {ha}).

>From the above discussion it is clear that, by following 
Ref.\cite {bak1}, one can  directly solve the eigenvalue problem for 
$\tilde {\cal H}_B $ (\ref {h1}) through 
  generalised Laguerre polynomials.  At present, however, 
we want to follow a different approach which attempts to 
solve this eigenvalue problem by mapping $\tilde {\cal H}_B $ 
 to a set of decoupled harmonic oscillators.
To this end, we notice that this 
effective Hamiltonian (\ref {h1})  may be expressed as
$ 
\tilde {\cal H}_B 
 = S^{-} + \omega S^3 $ , where the $B_N$ type Lassalle operator 
$S^{-}$ and the Euler operator $S^3$ are given by \cite {wad0}
\begin{eqnarray}
S^{-} &= &- \frac{ 1}{2}
\sum_{j=1} ^N \frac{ \partial ^2}{\partial x_j^2} 
 -2\tilde{\rho}
\sum_{\stackrel{j,k=1}{j\neq k}}^N 
\frac{ 1}{x_j^2 - x_k^2} \left ( x_j \frac{ \partial
}{\partial x_j} -x_k \frac{ \partial }{ \partial x_k}\right ) -\tilde{\sigma}
\sum_{j=1}^N \frac{ 1}{x_j}\frac{ \partial }{\partial x_j} \nonumber
\\ 
S^3 &=& \sum x_j \frac{ \partial }{ \partial x_j} .
\label{s3}
\end{eqnarray}
It is easy to see that the above defined operators satisfy the simple
 commutation relation: $ [S^3, \ S^{-} ] = -2 S^{-}$.  By
taking advantage of such commutation relation,
we perform further similarity transformations to  
$\tilde {\cal H}_B $
 and reduce it
to the Hamiltonian corresponding to free oscillators as
\begin{equation}
{\cal H}_{free}
 = {\cal S}^{-1} \tilde {\cal H}_B {\cal S}
 = - \frac{ 1}{2}\sum_{j=1}^N \frac{ \partial ^2}{
\partial x_j^2}+ \frac{ 1}{2} \omega \sum_{j=1}^N x_j^2 - \frac{ \omega
N}{2} 
\label{h2}
\end{equation}
where $ {\cal S}= {\cal S}_0  \, e^{ \frac{ \omega}{2}\sum x_i^2} $ and
 $ {\cal S}_0 = 
e^{ \frac{ 1}{2 \omega }S^-} e^{ \frac{ 1}{4 \omega}
   \sum_{j=1}^N \frac{ \partial ^2}{\partial x_j^2} } $.

Due to similarity transformations in (\ref {h1}) and (\ref{h2}),
one may naively think that the eigenfunctions of the extended 
$B_N$ Calogero model (\ref {ha}) can be obtained from those of 
the free oscillators as:
$\psi_{n_1,n_2,....., n_N} = \psi_{gr} {\cal S}_0 \Big (
H_{n_1}(x_1) H_{n_2}(x_2)\cdots H_{n_N}(x_n) \Big ) $,
where $H_{n_j}(x_j)$ denotes the Hermite polynomial of order $n_j$. 
However,  similar to the case of $B_N$ Calogero model \cite {gp,wad0},
the action of ${\cal S}_0$ leads to a singularity unless 
all $n_j$s are chosen to be even integers and 
$H_{n_1}(x_1) H_{n_2}(x_2)\cdots H_{n_N}(x_n) $  is symmetrised 
with respect to all coordinates. Therefore, nonsingular eigenfunctions of 
$\tilde {\cal H}_B $ 
(\ref {ha}) can be obtained from the eigenfunctions 
 of free oscillators as
\begin{equation}
\psi_{n_1,n_2,....., n_N} = \psi_{gr}  \, {\cal S}_0
 \Big ( \, \Lambda_{+} \, \left ( \,
H_{2n_1}(x_1) H_{2n_2}(x_2)\cdots H_{2n_N}(x_n)  \, \right ) \, \Big ) \, ,
\label{psi} 
\end{equation}
where 
$ \Lambda_{+} $ completely symmetrises all coordinates and thus 
projects  the  distinguishable many-particle  wave functions 
 to the bosonic part of the Hilbert space.
Eigenvalues of ${\cal H}_B$
corresponding to eigenfunctions (\ref {psi}) are given by
\begin{equation}
E_{n_1, n_2, ... n_N} = E_{gr} + 2 \omega \sum_{j=1}^N n_j =
\frac{ \omega N}{2} + \omega N \tilde{ \sigma }+
 \omega N(N-1)2\tilde{\rho} + 2 \omega \sum_{j=1}^N n_j ,
\label{eng}
\end{equation}
where the excitation numbers $n_j$s are non-negative integers obeying 
bosonic selection rule  $n_{j+1} \geq n_j $. 

Since 
$ \tilde{ \sigma }$ and $\tilde{\rho} $ are real parameters,
the energy eigenvalues (\ref {eng}) are also real ones.
Thus we interestingly find that 
 the PT invariant nonhermitian Hamiltonian (\ref {ha}) generates
real spectrum. However, within the above mentioned approach, 
it is not clear whether the corresponding eigenfunctions form a
complete set. 
It is also evident that, apart form a constant shift for all energy levels,
the spectrum (\ref{eng}) coincides with the 
spectrum of N number of free bosonic
oscillators with frequency $2 \omega $. As the term
  $2 \omega \sum_{j=1}^N n_j$  in eqn.(\ref{eng}) is always
non-negative, $E_{n_1, n_2, ... n_N}$ can not be less than $E_{gr}$.
Consequently, $\psi_{gr} $  (\ref {gr}) 
indeed  represents the ground state wave function
of ${\cal H}_B$ (\ref {ha}) with energy $E_{gr}$. 

Finally we want to point out an important difference between
the ground state energies of 
 $B_N$ Calogero model (\ref{hb})
 and its $PT$ invariant extension (\ref {ha}).
At the limit $\delta_1 = \delta_2 =0$,
 eqn.(\ref {egr}) reproduces the ground state energy of 
 $B_N$ Calogero model as
\begin{equation}
E_{gr} = \frac{ \omega N}{2} + \omega N  \sigma + \omega
N(N-1)2 \rho ,
\label{egr0}
\end{equation}
and the form of corresponding 
 ground state eigenfunction is given by (\ref {gr}).
 Since  $\sigma $ and $\rho $ must be positive 
 to ensure the nonsingularity of ground state eigenfunction (\ref {gr}),
 the ground state energy (\ref {egr0}) of the  $B_N$ Calogero model 
is always a positive quantity. However, 
as evident from eqn.(\ref {egr}), the 
 ground state energy of
the extended $B_N$ Calogero model is dominated by the $N^2$ order term 
for large values of $N$. The coefficient of this $N^2$ order term, i.e. 
  $2\tilde { \rho}$, 
is related to the coupling constants $ \delta_2  , \   g_2  $ through
a quadratic equation given by
\begin{equation}
  g_2 = (2 \tilde { \rho })^2 - 2 \tilde { \rho } - \delta _2  ( 1+ 
 \delta _2  ) \, .
\label{new}
\end{equation}
By using (\ref{new}) and the second relation of 
(\ref {g1g2}),  it is easy to see that 
   the parameter
 $2\tilde { \rho}$ becomes  negative (even though 
 $2 \rho $ remains positive)  if 
the coupling constants $ \delta_2  , \   g_2  $ are chosen
within the range: $ \delta_2 > 0 ,  \ 0 \geq  g_2 > - \delta_2 
(1+ \delta_2 )$.  
Therefore, for large values 
of $N$, the ground state energy of extended $B_N$ Calogero model (\ref {ha})
will also be a negative quantity within the above mentioned range of two
coupling constants.

\section{Some properties of extended $B_N$ Calogero model}

\subsection{Connection with fractional statistics}

Generalised exclusion statistics (GES)  introduced  by Haldane
\cite{hal} is believed to play an important role in the edge excitations 
of fractional quantum Hall effect.  Such exclusion statistics can be 
realised microscopically in 
usual Calogero models with hermitian Hamiltonians \cite {ms,ms1,ms2}.
The GES parameter for these Calogero models
 is a measure of `level repulsion' of the
quantum numbers generalising the Pauli exclusion principle \cite {ms2}.
So the partition function and various 
thermodynamical quantities for these Calogero models can be 
derived as a function of the corresponding GES parameters.
Now for exploring GES in the case of our PT invariant $B_N$ type
Calogero model (\ref {ha}), 
we observe that  eqn.(\ref{eng}) can be rewritten 
exactly in the form of energy spectrum for $N$ free oscillators as
\begin{equation}
E_{n_1,n_2\cdots n_N} = \frac{\bar{\omega}N}{2} + \bar{\omega}
\sum_{j=1}^N \bar{n_j} \, ,
\label{frac}
\end{equation}
where $\bar{\omega} = 2\omega$ and 
\begin{equation}
\bar{n_j} = n_j + 2\tilde{\rho} j + \frac{1}{4} ( 2\tilde{\sigma}-
8\tilde{\rho} -1)
\label {qua}
\end{equation}
are quasi-excitation numbers. However, from eqn.(\ref {qua}) it is evident
that such quasi-excitation numbers are no longer integers
and they satisfy a modified selection rule:
$ \bar{n}_{j+1}- \bar{n}_j \geq 2 \tilde{\rho} \,  .$ Thus the minimum 
 difference between two consecutive 
$\bar{n}_j $s is given by
\begin{equation}
2 \tilde{\rho} = 2\rho - \delta_2 \, ,
\label {selec}
\end{equation}
(here we assume $2\rho \geq \delta_2$).
As a consequence the spectrum of extended $B_N$ Calogero model (\ref {ha})
satisfy GES with parameter $2 \tilde{\rho}$. It is interesting to 
notice that this exclusion statistics parameter depends only two 
 coupling constants $\delta_2 $  and $g_2 $, which control the strength of 
long-range interactions in the Hamiltonian (\ref{ha}).
 Moreover, eqn.(\ref {new}) 
describe a parabolic curve on the $( \delta_2 , \ g_2) $ plane for any
fixed value of $2 \tilde{\rho}$.  Consequently, all points on such a
parabolic curve,   representing 
 extended $B_N$ Calogero models associated with different values of 
 $\delta_2 $  and $g_2 $, yield the same 
 exclusion statistics parameter.

It may be observed that
 the eigenfunctions (\ref {gr}) and (\ref {psi})  pick up
a phase factor $(-1)^{2 \rho }$ under the exchange of any two 
particles.  So the  exchange statistics  parameter for 
 the extended $B_N$ Calogero model (\ref {ha}) is given by $2\rho $.
It is clear from eqn.(\ref {selec}) that, for the case
$\delta_2 \neq 0$, the exchange statistics parameter for the 
  extended $B_N$ Calogero model differs from the 
corresponding exclusion statistics parameter.
Though the 
 the exchange statistics parameter is not directly related to the spectrum or
thermodynamics of the system, it has some interesting features.
For example,  in absence of confining harmonic potential,  
  the scattering phase shift 
for multi-particle scattering  generally depends 
 on the exchange statistics parameter \cite {ms2}.
  Moreover, quite similar 
to the case of usual Calogero models \cite {ms2},  
the exchange statistics parameter for the extended $B_N$ Calogero model 
fixes the boundary conditions on the wave functions 
 (\ref {gr}) and (\ref {psi}) at the limits $x_i \rightarrow x_j$.

\subsection{Reflection symmetry of the exchange statistics parameter}

It is well known that, 
for  fixed  values of all coupling constants,
the exchange statistics parameter of the $B_N$ Calogero  model (\ref {hb})
may take two distinct values.  These two distinct values of the 
 exchange statistics parameter  are also related through a 
 `reflection symmetry' given by: $2\rho \rightarrow 1 - 2\rho $.
  For exploring such reflection symmetry 
in the case of  extended $B_N$ Calogero model (\ref {ha}),
we note that the second  relation in (\ref{g1g2}) 
can be easily solved to obtain  two solutions of 
 the exchange statistics parameter as
\begin{equation} 
2\rho = \frac{1}{2}(1+2\delta_2)\pm \frac{1}{2}\sqrt{(1+2\delta_2)^2+4g_2}.
\label {ref}
\end{equation} 
Due to the existence of these two solutions, it appears that 
reflection symmetry is also present 
in the case of  extended $B_N$ Calogero model (\ref {ha}).
However, we have already noticed that  only  positive solutions
of the parameter $2\rho$ lead to physically acceptable
 nonsingular eigenfunctions of ${\cal H}_B$. So,
the reflection symmetry can exist in the case of 
 extended $B_N$ Calogero model provided both solutions of $2 \rho $ 
in eqn.(\ref {ref}) take non-negative values. It is easy to see that 
both of these solution will be non-negative only  within a parameter
range given by $(1+2\delta_2) > 0$ and $g_2 \leq 0$. 
Thus we curiously find 
that a kind of `phase transition' occurs at the line $\delta_2 =
- {1 \over 2}$ on the $(\delta_2 , g_2)$ plane. 
For the  case
 $\delta_2 \leq - {1 \over 2}$, the 
 reflection symmetry of the exchange statistics 
parameter is lost for  any  possible value of $g_2$.
 On the other hand, 
for the case $\delta_2 > - {1 \over 2}$
the exchange statistics parameter 
shows a reflection symmetry: $2\rho \rightarrow 1 + 2\delta_2
- 2\rho  \ $, if $g_2$ is chosen  within an interval 
$ - ( {1\over 2} + \delta_2 )^2 \leq g_2 \leq 0$.

\subsection{Relation with $B_N$ Calogero model}

We have seen in sec.II that, similar to the case of $B_N$ Calogero model, 
the extended $B_N$ Calogero model (\ref {ha}) can also be mapped
to a system of free harmonic oscillators through a similarity 
transformation. So it is natural to enquire whether 
the extended $B_N$ Calogero model (\ref {ha}) is directly 
related to the 
 $B_N$ Calogero model (\ref {hb}) through some similarity transformation.
Investigating along this line, we find that 
\begin{equation}
\Gamma ^{-1}{\cal H}_B \Gamma =  H_B^\prime 
 \equiv \frac{ 1}{2}\sum p_j^2 + \frac{ 1}{2}\omega ^2
\sum x_j^2 + \frac{ g_1^\prime }{2}\sum \frac{ 1}{x_j^2} + g_2^\prime 
\sum_{\stackrel {j,k=1}{j\neq k}}^N \frac{ x_j^2 + x_k^2}{x_j^2-x_k^2} \, ,
\label{sim}
\end{equation}
where  
\begin{equation}
\Gamma = \prod_{\stackrel{j,k=1}{j\neq k}}^N
 (x_j^2-x_k^2)^{ \frac{ \delta _2}{2}}\prod
_{l=1}^N x_l^{ \delta _1 },  
\label{gam}
\end{equation}
and $H_B^\prime $ denotes the Hamiltonian of $B_N$ 
Calogero model with  `renormalised' coupling constants 
 given by
\begin{equation}
  g_1^\prime = g_1 + \delta _1(1+\delta
_1), ~~~~ g_2^\prime = g_2 + \delta _2(1+\delta _2).
\label{re}
\end{equation}

However it should be observed that,
for any nonzero value of $ \delta_2 $,
either $\Gamma$ (\ref {gam}) or its inverse 
becomes singular at the limit $x_j \rightarrow x_k$.
So the relation (\ref {sim}) can not be interpreted as a similarity 
transformation in the usual sense and it may lead to some 
strange consequences. For example, due to  relation (\ref {sim}), one may
expect that the Hamiltonians 
  ${\cal H}_B$ and $H_B^\prime$  must share exactly same eigenvalues.
However,  we have already noticed in sec.II that the ground 
state energy of $H_B^\prime$ is always a positive quantity, while 
 the ground state energy of ${\cal H}_B$ (\ref {ha}) 
can be a negative quantity provided the  exclusion 
statistics parameter $ 2 \tilde {\rho } $ takes a negative value. 
 To explain this  rather unexpected result, we first recall that 
 $2 \tilde {\rho }$ will be negative if the coupling
constants of ${\cal H}_B$ (\ref {ha}) are chosen within the
range:    $\delta_2 >0 $ and 
$ 0 \geq  g_2 > -  \delta_2 ( 1+ \delta_2 )$. So, from eqn.(\ref {re})
one finds that  the renormalised coupling constant 
$ g_2^\prime $ must be a positive quantity in this case.
Consequently, the corresponding exclusion statistics parameter 
$2\rho^\prime $, which is obtained by solving the 
 quadratic equation
 $ g_2^\prime  = 2 \rho^\prime  ( 2\rho^\prime -1) $,  has 
one positive  and one negative solution. The positive solution
of $2\rho^\prime $  leads to a 
ground state for  $H_B^\prime $ with positive eigenvalue. 
 On the other hand, for large values of $N$, the negative solution of 
$2\rho^\prime $ yields a ground state for  $H_B^\prime $ with negative 
eigenvalue.  However, one usually throws away this negative solution of 
 $ 2 \rho^\prime $, since the corresponding ground state eigenfunction 
$ \psi_{gr}^\prime (x_1, x_2, \cdots , x_N )$ becomes 
singular at the limit $x_j \rightarrow x_k$.  Nevertheless, as
 will be shown below,  this unphysical eigenfunction 
of $H_B^\prime $ enables us to construct
a physical eigenfunction of ${\cal H}_B$.
Due to the relation (\ref {sim}), one finds that
\begin{equation}
\psi_{gr} (x_1, x_2, \cdots , x_N ) ~=~ 
 \Gamma \psi^\prime_{gr} (x_1, x_2, \cdots , x_N ) \,  ,
\label{wave}
\end{equation}
where 
$ \psi_{gr} (x_1, x_2, \cdots , x_N )$ represents 
the  ground state eigenfunction of 
${\cal H}_B$.
It can be easily checked that 
 the zeros of $ \Gamma$ are sufficient to cancel all singularities of 
 $ \psi_{gr}^\prime (x_1, x_2, \cdots , x_N )$ 
 at the limit $x_j \rightarrow x_k \ $. Therefore, the ground state  
 eigenfunction $\psi_{gr} (x_1, x_2, \cdots , x_N ) $ (\ref {wave})
 is no longer singular at this limit.  Thus we curiously find that 
a singular eigenfunction of $H_B^\prime $ generates a nonsingular 
  eigenfunction of ${\cal H}_B$  through the transformation
(\ref {wave}).  Consequently, the ground state energy of 
${\cal H}_B$ (\ref {ha})  coincides with the  negative 
eigenvalue of $H_B^\prime $ associated with its unphysical eigenfunction
 $ \psi_{gr}^\prime (x_1, x_2, \cdots , x_N )$. In a similar way one can 
show that all singular 
eigenfunctions of $H_B^\prime $, which represent
the excited states over the unphysical ground state eigenfunction
 $ \psi_{gr}^\prime (x_1, x_2, \cdots , x_N )$, generate 
nonsingular excited states of 
${\cal H}_B$ (\ref {ha}) through a transformation by $\Gamma $.

\section{Conclusion}

Here we observe that
a recently considered nonhermitian variant of 
$A_{N-1}$ Calogero Hamiltonian with real spectrum remains invariant 
under the PT transformation.  Being encouraged 
by this observation,  we propose a new many-particle
quantum system (\ref {ha}) by adding some nonhermitian 
but PT invariant interactions to the $B_N$ Calogero model.
Such PT invariant interactions depend on both coordinate and
momentum variables of the particles.
By using appropriate similarity transformations, we are able 
to map this extended 
 $B_N$ Calogero model to a set of free harmonic oscillators
and solve this model exactly. It turns out that this 
many-particle system with nonhermitian Hamiltonian 
yields a real spectrum. This 
fact supports the conjecture that
  the condition of hermiticity 
on a Hamiltonian can be replaced by the weaker condition of PT symmetry 
to ensure that the corresponding  eigenvalues would be real ones. 
It is also found that the spectrum of extended 
 $B_N$ Calogero model obeys a selection rule
which leads to generalised 
exclusion statistics (GES). 
However, the extended
 $B_N$ Calogero model also possess some remarkable properties 
which are absent in the case of usual
 $B_N$ Calogero model. For example, we curiously find that the
 GES parameter for this extended $B_N$ Calogero model 
 differs from the corresponding exchange statistics parameter. Moreover
a `reflection symmetry' of the 
  exchange statistics parameter, which  is known to exist for 
   $B_N$ Calogero model, can  be found
 in the case of extended model only if  
 the strength of a PT invariant interaction 
is chosen above a critical value. 
  As a future study, it will 
be interesting to search for other exactly solvable many-particle 
systems with nonhermitian but PT invariant Hamiltonians and 
investigate the nature of their eigenvalues as well as eigenfunctions.

\vspace {.5cm}
\noindent{\bf Acknowledgements}
\\
We like to thank the Referee for many constructive comments which helped to 
improve the presentation of this paper.

\end{document}